\documentclass[pre,twocolumn,aps,prb,floatfix]{revtex4-2}

\usepackage{amsmath,amssymb,amsfonts}
\usepackage{graphicx}
\usepackage{bm}
\usepackage{booktabs}
\usepackage{braket}
\usepackage{xcolor}
\usepackage[normalem]{ulem}

\usepackage{orcidlink}
\begin{document}

\title{Multi-scale spectral statistics of intermediate quantum chaos in ultracold erbium collisions}

\author{Iv\'an R. R. Gonz\'alez \orcidlink{0000-0003-1853-3748}}
\email{ivan.roa@umayor.cl}
\affiliation{School of Engineering, Faculty of Science, Engineering and Technology, Universidad Mayor, Santiago 7500994, Chile}
\affiliation{Centro Multidisciplinario de Física, Universidad Mayor, Camino la Pirámide 5750, Huechuraba, Santiago, Chile}

\author{Barnali Chakrabarti \orcidlink{0000-0002-6320-9894}}
\email{barnali@if.usp.br}

\affiliation{Laboratorio de F\'isica Teórica e Computacional, Departamento de F\'isica, Universidade Federal de Pernambuco, 50670-901 Recife, Pernambuco, Brazil}

\author{Ant\^onio M. S. Mac\^edo \orcidlink{0000-0002-4522-031X}}
\email{antonio.smacedo@ufpe.br}
\affiliation{Laboratorio de F\'isica Teórica e Computacional, Departamento de F\'isica, Universidade Federal de Pernambuco, 50670-901 Recife, Pernambuco, Brazil}

\date{\today}

\begin{abstract}
Characterizing intermediate quantum chaos in systems with mixed phase spaces requires probing spectral correlations across multiple scales. 
Here, we present a multi-scale statistical analysis of the resonance spectra of ultracold $^{166}$Er and $^{168}$Er atoms using higher-order spacing ratios, spacing increments, and the accumulated power spectral density of decimated level sequences. Our analysis reveals that while the spectra exhibit clear level repulsion, they are consistently better described by semi-Poisson than by standard Wigner–Dyson ensembles, with this intermediate character persisting across higher-order correlations. The accumulated power spectrum reveals an intermediate $1/f^\alpha$ regime, with $\alpha\simeq1.5$ at full spectral resolution and a tendency toward the integrable random-walk limit $\alpha=2$ at intermediate decimation scales, indicating a scale-dependent loss of spectral rigidity. Furthermore, the spacing-increment distributions independently reveal statistics intermediate between Laplace and Gaussian limits. Together, these complementary diagnostics show that the semi-Poisson character reflects persistent non-Wigner–Dyson correlations beyond the nearest-neighbor scale, providing a consistent picture of intermediate quantum chaos and  a possible  connection to hierarchical trapping in the underlying collision dynamics.

\end{abstract}

\maketitle

\section{Introduction}
\label{sec:introduction}

The statistical analysis of quantum energy spectra provides a powerful framework for characterizing underlying dynamical regimes, with the Berry-Tabor and Bohigas-Giannoni-Schmit (BGS) conjectures establishing the two canonical limits of integrable and fully chaotic quantum dynamics. According to the Berry-Tabor conjecture, generic integrable systems exhibit uncorrelated spectral fluctuations, leading to a Poissonian nearest-neighbor spacing distribution (NNSD), $P(s)$~\cite{Berry1977}. In contrast, the BGS conjecture associates quantum systems whose classical counterparts are fully chaotic with the universal spectral correlations of random-matrix theory (RMT), manifested through Wigner-Dyson statistics~\cite{Bohigas1984,Weidenmuller2009}. This integrable-chaotic dichotomy is commonly characterized using local spectral measures, such as the NNSD and ratios of consecutive level spacings, $r_n$, which capture level clustering in the integrable limit and level repulsion in the chaotic regime~\cite{Atas2013, Haake2010}.

However, many complex quantum systems do not conform strictly to these asymptotic limits. Systems with mixed phase-space dynamics--where a chaotic sea is interwoven with regular islands and hierarchical structures~\cite{Robnik1986,Meiss1992}--can exhibit intermediate quantum chaos, characterized by spectral statistics intermediate between Poisson and Wigner-Dyson limits, including semi-Poisson behavior~\cite{Bogomolny1999}. Diagnosing this intermediate regime requires moving beyond strictly local spectral diagnostics. Moreover, the extraction of the nearest-neighbor spacing distribution $P(s)$ from finite experimental spectra is sensitive to the unfolding procedure, which requires an estimate of the local mean level density and can introduce spurious correlations. The ratio statistic $r_n$, introduced by Atas \textit{et al.}~\cite{Atas2013}, largely circumvents this unfolding ambiguity and has therefore become a robust tool for analyzing finite spectra. Nevertheless, conventional consecutive-spacing ratios remain intrinsically local and provide limited information about correlations extending beyond neighboring levels. This motivates the use of higher-order spacing ratios, which probe spectral correlations over progressively larger level separations and provide a natural bridge from local statistics toward mesoscopic spectral structure.

To achieve a comprehensive characterization of intermediate quantum chaos, it is necessary to probe spectral correlations across multiple scales and complementary representations. In this work, we apply three complementary diagnostics to a common experimental benchmark: the Fano-Feshbach resonance spectra of the heavy lanthanide isotopes $^{166}$Er and $^{168}$Er~\cite{Frisch2014}. These ultracold collision spectra provide a paradigmatic setting for investigating complex quantum dynamics in atomic systems. The large magnetic moments of erbium, together with strong spin-orbit coupling and anisotropic dipole-dipole interactions~\cite{Baumann2010}, lead to dense and strongly coupled molecular-state spectra, providing an experimentally accessible platform for studying deviations from the universal Poisson and Wigner-Dyson limits~\cite{Roy2017}.

 Previous analyses of these erbium spectra established their close-to-semi-Poisson character, but the spectral origin of these intermediate statistics remained unresolved~\cite{MurPetit2015,Roy2017}. In particular, it remained unclear whether the semi-Poisson behavior was confined to nearest-neighbor correlations or persisted across larger level separations. Addressing this question requires going beyond conventional local statistics and examining how spectral correlations evolve across multiple scales.

Rather than relying on a single spectral diagnostic, we examine three complementary observables that probe spectral correlations at different scales. First, we analyze \textit{higher-order spacing ratios}, $r^{(k)}$, which probe correlations between levels separated by progressively larger level intervals~\cite{Tekur2018}. By comparing their empirical distributions with the effective-index surmise of RMT and the semi-Poisson beta-prime family, we characterize the evolution of level correlations beyond the nearest-neighbor scale.
%
%
Second, to characterize \textit{longer-range spectral correlations}, we treat the ordered sequence of unfolded energy levels as a discrete signal. The power spectral density (PSD) of the accumulated spectral fluctuations connects spectral rigidity with colored-noise statistics: fully chaotic spectra are associated with $1/f$ noise ($\alpha=1$), whereas integrable spectra exhibit Brownian $1/f^2$ noise ($\alpha=2$)~\cite{Relano2002,Gomez2005}. We extend this approach to decimated level sequences and use the resulting accumulated PSD to investigate how spectral rigidity changes under systematic coarse-graining. This frequency-domain analysis provides a complementary probe of spectral correlations beyond the local and higher-order statistics and allows us to investigate signatures of hierarchical structure in the spectrum~\cite{Zaslavsky2000,Iomin2000}.

Third, we consider the statistics of \textit{spacing increments}, an additive measure of fluctuations between consecutive spacings. Comparing the increment distributions with Laplace and Gaussian reference forms provides an alternative characterization of the integrable-chaotic crossover in terms of spectral fluctuations~\cite{Gonzalez2025}. 
%

The paper is organized as follows. Section~II develops the theoretical framework for the higher-order spacing ratios and the decimated accumulated power spectral density. Section~III introduces the stochastic hierarchical framework and its application to the erbium resonances. Section~IV details the spacing-increment formalism and the integrable-chaotic crossover. Section~V discusses the implications of our multi-scale triangulation, and Section~VI summarizes our conclusions.

\section{Theoretical Framework for Spectral Fluctuations}\label{sec:theory}

\subsection{The ATAS Surmise and the Multiplicative Structure of Spectral Fluctuations}
\label{sec:atas}

Atas, Bogomolny, Giraud, and Roux (ATAS) introduced the ratio of consecutive spacings, $r_n = s_n / s_{n-1}$, as a dimensionless, scale-invariant probe that avoids the need for unfolding~\cite{Atas2013}.
Fundamentally, both the multiplicative spacing ratios and the additive spacing increments analyzed in subsequent sections are marginal projections of a single parent object: the joint probability density of two consecutive gaps, $P_c(s_1, s_2)$. For the ratio variable, the exact probability density function (PDF) is obtained via the integral projection
\begin{equation}
    P(r) = \int_0^\infty ds_1 \, s_1 \, P_c(s_1, r s_1).
\end{equation}
By evaluating this projection over the joint eigenvalue density of $3 \times 3$ Gaussian random matrices, ATAS derived a highly accurate Wigner-like surmise for the Gaussian $\beta$-ensemble:
\begin{equation}
    P_\beta(r) = \frac{1}{Z_\beta} \frac{(r + r^2)^\beta}{(1 + r + r^2)^{1 + \frac{3\beta}{2}}}, \quad r \ge 0,
    \label{eq:atas_surmise}
\end{equation}
where the Dyson index $\beta \in \{1, 2, 4\}$ corresponds to the Gaussian orthogonal (GOE), unitary (GUE), and symplectic (GSE) ensembles~\cite{Mehta2004}, respectively. The normalization constant $Z_\beta$ is analytically given by
\begin{equation}
    Z_\beta = \frac{2\pi \Gamma(1+\beta)}{3^{3(1+\beta)/2} \Gamma(1+\beta/2)^2}.
    \label{eq:Zbeta}
\end{equation}
While remarkably accurate for practical applications, it is crucial to note that Eq.~\eqref{eq:atas_surmise} is a finite-$N$ surmise. In the large-matrix limit ($N \to \infty$) of the unitary class, the bulk statistics are governed by a determinantal point process. The exact two-gap density $P_c(s_1, s_2)$ must then be constructed from one-point Janossy densities via $P_c(a_1, a_2) = -\partial_{a_1}\partial_{a_2} J_1(0; [a_1, a_2])$, where the Fredholm determinant is constrained by a two-endpoint Tracy-Widom differential system~\cite{Tracy1994,nishigaki2024,Nishigaki2021}. Following Nishigaki's notation, $a_1<0<a_2$ denote the
positions of the neighboring eigenvalues relative to an eigenvalue
conditioned to lie at the origin. The corresponding positive
spacings are $s_1=-a_1$ and $s_2=a_2$. We retain the symbol $P_c$
for both representations, distinguishing them by their arguments:
$P_c(s_1,s_2)=P_c(-s_1,s_2)$, where the left-hand side denotes
the density of positive spacings and the right-hand side denotes
the density of signed positions. The ATAS surmise effectively captures the core features of this exact integrable structure without the computational overhead associated with evaluating the corresponding Fredholm determinants and solving the associated Painlevé transcendental equations~\cite{nishigaki2024}.

The ATAS surmise is constrained by deep structural symmetries inherent to the random matrix ensembles. The most prominent of these is the inversion symmetry,
\begin{equation}
    P_\beta(r) = \frac{1}{r^2} P_\beta\left(\frac{1}{r}\right),
\end{equation}
which follows directly from the left-right symmetry of the joint gap density $P_c(s_1, s_2) = P_c(s_2, s_1)$. Physically, this relation reflects the statistical symmetry of the spectral staircase under the exchange of adjacent spacings. The Jacobian of the transformation $r \to 1/r$ yields the $1/r^2$ prefactor, ensuring invariance of the probability measure under this exchange. For stationary and reflection-symmetric spectra, this relation provides a useful consistency check on empirical ratio distributions.

Furthermore, the asymptotic limits of $P_\beta(r)$ encode critical information regarding  local level repulsion and relative fluctuations of adjacent spacings. In the limit $r \to 0$, the surmise exhibits a power-law repulsion, $P_\beta(r) \sim r^\beta$, which is the direct multiplicative analogue of the Wigner-Dyson level repulsion $P(s) \sim s^\beta$. It quantifies the suppression of very small level spacings dictated by the underlying symmetries (e.g., time-reversal and spin-rotation)~\cite{Altland1997}. Conversely, in the limit $r \to \infty$, the distribution decays algebraically as $P_\beta(r) \sim r^{-(2+\beta)}$. This heavy, power-law tail stands in stark contrast to the rapidly decaying large-spacing tail of the standard spacing distribution $P(s)$, capturing the rare but statistically significant events where a very small spacing is adjacent to a much larger gap.

To regularize the divergent moments associated with this algebraic tail and facilitate direct comparison with finite experimental histograms, it is standard practice to restrict the ratio to the compact interval $[0, 1]$ by defining the bounded ratio $\tilde{r}_n = \min(r_n, 1/r_n)$. The corresponding bounded probability density is simply $\widetilde{P}_\beta(\tilde{r}) = 2 P_\beta(\tilde{r}) \Theta(1 - \tilde{r})$. This formulation yields well-defined surmise averages, such as $\langle \tilde{r} \rangle_{\text{GOE}} \approx 0.536$, $\langle \tilde{r} \rangle_{\text{GUE}} \approx 0.603$, and $\langle \tilde{r} \rangle_{\text{GSE}} \approx 0.676$, which monotonically increase with the degree of level repulsion. 

In the specific context of the ultracold erbium collisions analyzed in this work, the ATAS surmise serves as the chaotic null hypothesis. However, the presence of the large magnetic moment of erbium, together with
strong spin-orbit coupling, and highly anisotropic dipole-dipole interactions leads to a complex coupled-channel molecular spectrum~\cite{Petrov2014} in which deviations from fully ergodic random-matrix behavior may arise. In such scenarios, hierarchical trapping near regular islands can reduce long-range spectral rigidity, driving the empirical distributions away from the strict Wigner-Dyson paradigm and toward intermediate statistics, such as semi-Poisson statistics. 

The nearest-neighbor ratio $r_n$ relies exclusively on the local correlations of three adjacent levels ($E_{n-1}, E_n, E_{n+1}$), and is therefore inherently limited in its sensitivity to mesoscopic spectral structures. This limitation necessitates the extension to higher-order spacing ratios, which probe the multiplicative correlations across increasingly distant energy scales.
\subsection{Higher-Order Spacing Ratios and the Effective Dyson Index}

While the nearest-neighbor ratio $r_n$ probes the local multiplicative structure of three adjacent levels, characterizing spectral correlations beyond the nearest-neighbor scale requires extending the analysis to mesoscopic energy scales. This is achieved by partitioning the spectrum into non-overlapping blocks of $k$ consecutive spacings. Defining the $k$-th order block spacing as $s_n^{(k)} = E_{n+k} - E_n = \sum_{j=0}^{k-1} s_{n+j}$, the higher-order spacing ratio is constructed as the quotient of adjacent blocks:
\begin{equation}
    r_n^{(k)} = \frac{s_{n+k}^{(k)}}{s_n^{(k)}} = \frac{E_{n+2k} - E_{n+k}}{E_{n+k} - E_n}.
\end{equation}
This decimation procedure effectively coarse-grains the spectral staircase, allowing spectral correlations to be examined at progressively larger level separations.

In the framework of random-matrix theory~\cite{Forrester2010}, the exact evaluation of the probability density $P^{(k)}(r)$ for $k > 1$ requires integrating the  joint density of $2k+1$ consecutive eigenvalues over $2k-2$ internal eigenvalues, a task that rapidly becomes analytically intractable. However, Tekur, Bhosale, and Santhanam demonstrated that the statistical properties of these block ratios are governed to remarkable accuracy by an \textit{effective-index rule}~\cite{Tekur2018}. The distribution $P^{(k)}(r; \beta)$ is exceptionally well approximated by the standard ATAS surmise evaluated at an effective Dyson index:
\begin{equation}
    \beta_{\text{eff}}(k, \beta) = \frac{k(k+1)}{2}\beta + (k - 1).
\end{equation}
This effective-index relation provides a compact description of the progressive narrowing of the higher-order ratio distribution with increasing $k$. For the Gaussian orthogonal ensemble ($\beta=1$), this yields the sequence $\beta_{\text{eff}} = [1, 4, 8, 13, 26, 43, 64, 89]$ for $k = [1, 2, 3, 4, 6, 8, 10, 12]$, respectively.

Crucially, the higher-order ratios inherit the fundamental structural symmetries of the nearest-neighbor case. Because the joint probability density of two adjacent $k$-blocks, $P_c^{(k)}(a, b)$, is symmetric under the exchange $a\leftrightarrow b$ for a stationary, reflection-symmetric spectrum,  the exact inversion symmetry is strictly preserved:
\begin{equation}
    P^{(k)}(r) = \frac{1}{r^2} P^{(k)}\left(\frac{1}{r}\right).
\end{equation}
This scale-free reversibility ensures that the bounded ratio $\tilde{r}^{(k)} = \min(r^{(k)}, 1/r^{(k)})$ remains a robust observable, free from artifacts associated with the ordering of the spectrum.

For increasing $k$, the effective-index ATAS distribution becomes progressively narrower around $r=1$. Thus, within the RMT description, higher-order ratios provide a sensitive probe of the persistence of long-range spectral correlations. However, this effective-index paradigm assumes that the long-range correlations characteristic of Wigner-Dyson ensembles persist across the energy scales probed by the ratio. In systems with mixed classical dynamics, such as the ultracold erbium collisions analyzed in this work, regular structures embedded in a chaotic background can modify these long-range correlations and produce intermediate spectral statistics.

This distinction is particularly important in the tails of the higher-order distributions. The ATAS surmise with $\beta_{\text{eff}}$ predicts a power-law tail
$P^{(k)}(r) \sim r^{-(2+\beta_{\text{eff}})}$, implying a rapidly decreasing probability of large block-ratio fluctuations as $k$ increases. In contrast, the semi-Poisson statistics~\cite{Gerland2001}, which incorporate short-range level repulsion without the full long-range rigidity of Wigner-Dyson spectra, yield a beta-prime family for higher-order ratios~\cite{Bogomolny2008,Akhshani2025}:
\begin{equation}
    P^{(k)}_{\text{SP}}(r) =
    \frac{\Gamma(4k)}{\Gamma(2k)^2}
    \frac{r^{2k-1}}{(1+r)^{4k}},
\end{equation}
whose tail decays as $r^{-(2k+1)}$. The comparison of the empirical erbium distributions with these two reference forms therefore provides a way to assess whether the higher-order spectral correlations follow the rigid RMT prediction or retain the weaker correlations characteristic of semi-Poisson statistics.

\subsection{Erbium Fano-Feshbach Spectra: Empirical Validation of Higher-Order Statistics}

The experimental benchmark for the higher-order ratio analysis is provided by the Fano-Feshbach resonance spectra of the heavy lanthanide isotopes $^{166}$Er ($N=189$ resonances) and $^{168}$Er ($N=190$ resonances), measured in the ultracold scattering experiments of Frisch \textit{et al.}~\cite{Frisch2014}. 
Comparable chaotic Fano-Feshbach resonance spectra have been reported for the neighboring highly magnetic lanthanide dysprosium~\cite{Maier2015}, indicating that this chaotic-resonance scenario also occurs in other open-shell lanthanide atoms.
These spectra constitute a paradigmatic realization of quantum chaos in atomic physics.
Following the protocol established by Roy \textit{et al.}~\cite{Roy2017}, we therefore treat the resonance positions as sharp eigenvalues, discarding width information to compare the real resonance positions with the spectral statistics of Hermitian RMT ensembles. The sequences are rescaled by their global mean spacing so that $\langle s \rangle = 1$, a choice that reduces the risk of the introduction of spurious correlations from higher-order polynomial fits, which are particularly problematic for short sequences of length $N \sim 190$.
The higher-order ratios $r^{(k)}_n = (E_{n+2k} - E_{n+k}) / (E_{n+k} - E_n)$ are computed for decimation orders $k \in \{1, 2, 3, 4, 6, 8, 10, 12\}$, with empirical histograms constructed using 35 uniform bins over the interval $[0, 5]$. These distributions are juxtaposed against two competing theoretical references: the effective-index ATAS surmise evaluated at $\beta_{\mathrm{eff}}(k,1)$, representing the Wigner–Dyson prediction for higher-order correlations, and the semi-Poisson beta-prime family, which models short-range repulsion without the long-range rigidity characteristic of Wigner–Dyson spectra. To ensure numerical stability at large $k$, all normalization constants involving products of gamma functions are evaluated in logarithmic space via $\ln( \Gamma)$.

\begin{figure*}[t!]
\includegraphics[width=\linewidth]{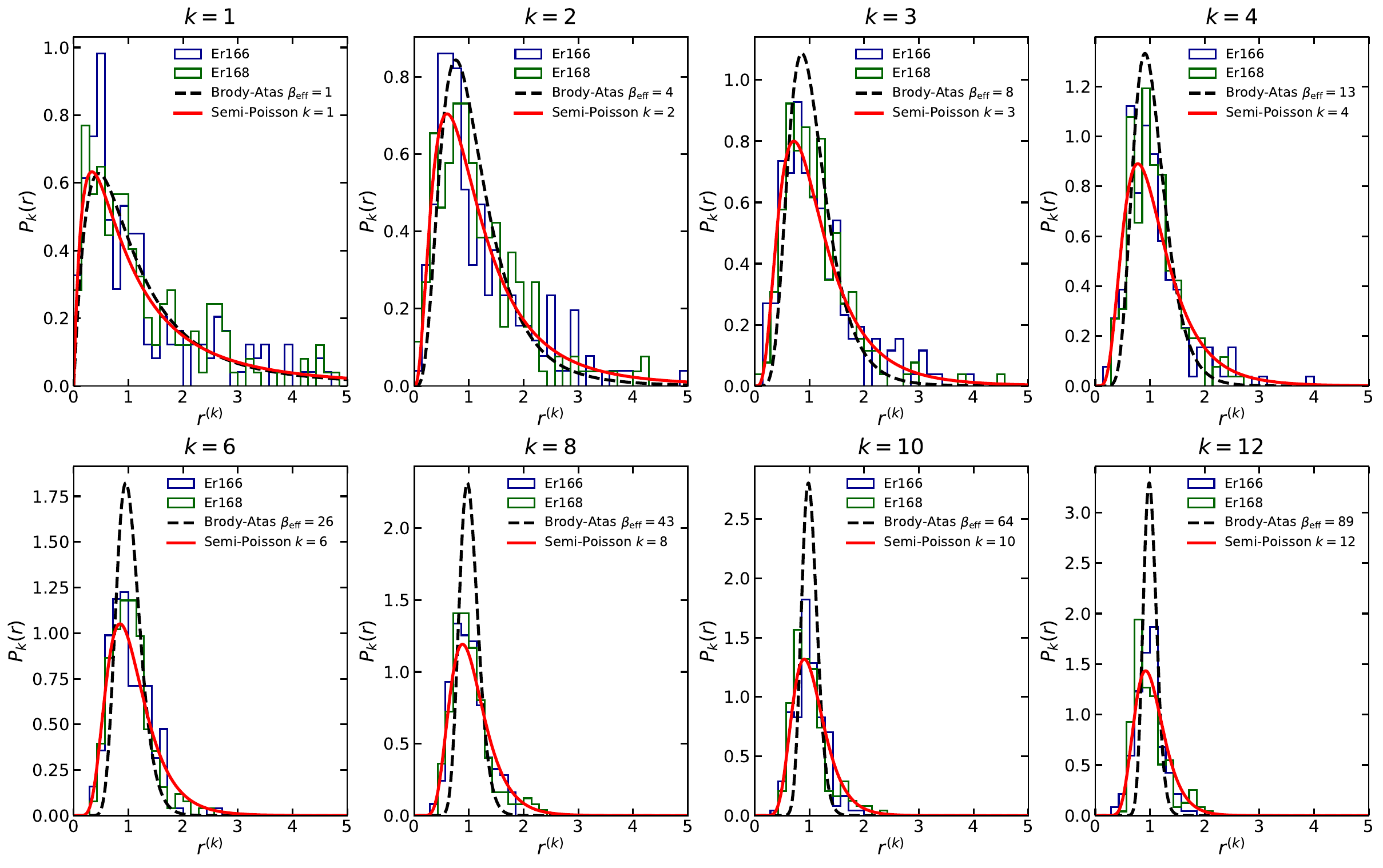}
\caption{Higher-order spacing-ratio distributions $P_k(r)$ for $k=1,2,3,4,6,8,10,12$.  Step histograms: empirical data for $^{166}$Er and $^{168}$Er.  Dash-dotted: ATAS surmise at $\beta_{\mathrm{eff}}$.  Dotted: semi-Poisson beta-prime law.}
\label{fig:ratios}
\end{figure*}

Figure~\ref{fig:ratios} displays the empirical distributions of $r^{(k)}$ for $k=1,\ldots,12$, compared against the ATAS surmise at $\beta_{\mathrm{eff}}$ and the semi-Poisson beta-prime law.  At $k=1$ the data lie between Poisson and GOE with the tails showing greater affinity with the semi-Poisson form, consistent with the findings of Ref.~\cite{Roy2017}.

As $k$ increases, the histograms become narrower, whereas the effective-index ATAS curve becomes increasingly concentrated around $r=1$ and progressively narrower than the experimental distributions. By contrast, the semi-Poisson reference provides a more accurate description of the data across the investigated values of $k$.

The higher-order ratio analysis demonstrates that the intermediate spectral character of the erbium Fano-Feshbach resonances persists beyond nearest-neighbor correlations. Although the spectra exhibit clear level repulsion, their higher-order ratio distributions remain systematically closer to the semi-Poisson beta-prime family than to the Wigner-Dyson/ATAS predictions over the investigated decimation orders. Moreover, the experimental distributions do not follow the progressively narrower effective-index RMT prediction as the level separation increases, indicating that the stronger correlations characteristic of Wigner-Dyson statistics do not fully develop over the accessible spectral scales. These results provide evidence for a persistent intermediate spectral regime in which local level repulsion coexists with weaker higher-order correlations than those predicted by Wigner-Dyson statistics.

These observations provide an important insight into the origin of the semi-Poisson character observed in the erbium spectra. Rather than being restricted to nearest-neighbor level repulsion, the intermediate statistics persist systematically at larger level separations, revealing persistent non-Wigner-Dyson correlations in the spectrum. Thus, the observed semi-Poisson behavior reflects a broader multiscale organization of spectral fluctuations, in which short-range level repulsion is accompanied by weaker higher-order correlations. This provides a spectral basis for the intermediate statistics and motivates the complementary analysis of spacing increments and accumulated power spectra presented below.

\subsection{Decimated Accumulated Power Spectral Density}

\begin{figure*}[t!]
\includegraphics[width=\linewidth]{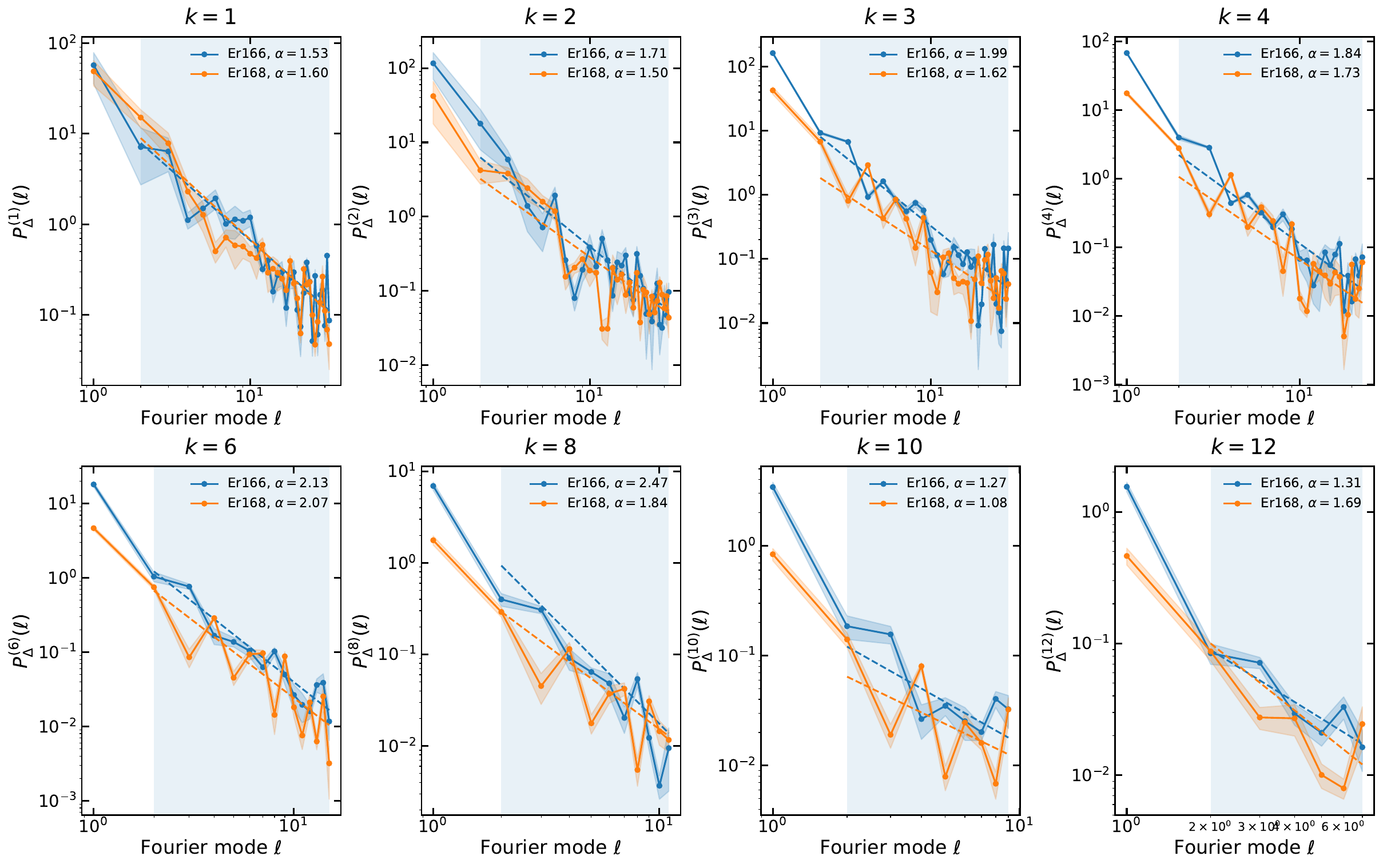}
\caption{Decimated accumulated PSD for $k=1,\ldots,12$.  The drift of
$\alpha$ toward 2 at intermediate $k$ signals the progressive loss of
spectral rigidity at mesoscopic scales.}
\label{fig:psd_grid}
\end{figure*}

While spacing ratios and increments probe the local and mesoscopic structure of spectral fluctuations in the energy domain, the global rigidity of the spectrum is most naturally characterized in the frequency domain. By treating the ordered sequence of unfolded energy levels as a discrete signal, one can map the concept of spectral rigidity directly onto the theory of colored noise. 

For a standard unfolded spectrum $\{E_n\}$ with unit mean spacing, the deviation of the $n$-th level from its mean position is given by the accumulated fluctuation $\Delta_n = E_n - E_0 - n$. The power spectral density (PSD) of this sequence, defined as the squared modulus of its discrete Fourier transform, provides a direct measure of the spectral form factor. A fundamental dichotomy emerges in the low-frequency behavior of the PSD, which follows a power law $S(f) \sim 1/f^\alpha$. According to the universal predictions of random-matrix theory, fully chaotic systems with strong long-range rigidity exhibit $1/f$ noise ($\alpha = 1$), whereas integrable systems, lacking long-range correlations, display Brownian $1/f^2$ noise ($\alpha = 2$).

To investigate how spectral rigidity evolves across different energy scales, we generalize this framework to decimated sequences. For a given decimation order $k$ and residue class $a \in \{0, \dots, k-1\}$, we extract the subsequence of levels $E^{(k,a)}_m = (E_{a+mk} - E_a)/k$. The local mean spacing within each decimated subsequence is renormalized to unity, and the corresponding accumulated fluctuation is computed as
\begin{equation}
    \Delta^{(k,a)}_m = \sum_{j=0}^{m-1} \left( S^{(k,a)}_j - 1 \right) = E^{(k,a)}_m - E^{(k,a)}_0 - m,
\end{equation}
where $S^{(k,a)}_j = E^{(k,a)}_{j+1} - E^{(k,a)}_j$ are the decimated spacings. The decimated accumulated PSD is then obtained by averaging the periodograms over all residue classes and available overlapping windows:
\begin{equation}
    P^{(k)}_\Delta(\ell) = \left\langle \left| \frac{1}{\sqrt{M}} \sum_{m=0}^{M-1} \Delta^{(k,a)}_m e^{-2\pi i \ell m / M} \right|^2 \right\rangle_{a, \text{win}},
\end{equation}
where $M$ is the number of points in the window and $\ell$ denotes the Fourier mode. The scale-dependent rigidity is quantified by fitting the ensemble-averaged PSD to the power-law model $\langle P^{(k)}_\Delta(\ell) \rangle \simeq C_{\beta,k} \ell^{-\alpha_{\beta,k}}$ over the selected frequency range.

The physical significance of the decimated PSD lies in its ability to probe how spectral correlations evolve from microscopic to mesoscopic scales. At $k=1$, the exponent $\alpha$ characterizes the rigidity of the full spectrum. As $k$ increases, the decimation procedure systematically coarse-grains the spectral staircase, suppressing sensitivity to the shortest-range fluctuations and emphasizing correlations over progressively larger level separations. In systems with mixed-phase-space dynamics, where hierarchical trapping can weaken long-range spectral rigidity, the exponent is expected to move away from the chaotic value $\alpha=1$ toward the integrable random-walk limit $\alpha=2$ as the decimation scale increases. The resulting scale dependence therefore provides a frequency-domain probe of spectral rigidity across scales that complements the local and higher-order statistics.

Given the finite length of the experimental sequences ($N\approx190$), the periodograms at large $k$ are inherently short and increasingly susceptible to statistical fluctuations and high-frequency outliers. To extract the spectral exponent $\alpha$ robustly, we employ the Theil-Sen estimator~\cite{Theil1950,Sen1968}, which computes the median of all pairwise slopes in the log-log plane over the fit window $2\leq\ell\leq M/2$. This nonparametric estimator is less sensitive to outliers than ordinary least-squares regression and does not require Gaussian error assumptions. Nevertheless, it does not eliminate finite-size effects, which become increasingly important at large decimation orders. We therefore interpret the large-$k$ values primarily as evidence of the scale dependence of the spectral fluctuations, rather than as precise estimates of an asymptotic exponent.


Figure~\ref{fig:psd_grid} extends the PSD analysis to all investigated values of $k$, with the resulting exponents collected in Table~\ref{tab:alpha}. For $^{166}\mathrm{Er}$, the exponent increases from $\alpha=1.53$ at $k=1$ to values near and above the Brownian limit, reaching $\alpha=2.47$ at $k=8$, before decreasing to $\alpha\simeq1.3$ at $k=12$, where only seven modes are available. For $^{168}\mathrm{Er}$, a similar increase toward the Brownian regime is observed at intermediate decimation orders, with $\alpha$ rising from $1.60$ at $k=1$ to $2.07$ at $k=6$, followed by substantial fluctuations at larger $k$. The intermediate-$k$ enhancement of $\alpha$ toward the integrable value indicates a reduction of spectral rigidity under coarse-graining, while the pronounced large-$k$ scatter is consistent with the increasing finite-size limitations of the decimated sequences. Thus, rather than exhibiting a monotonic flow from chaotic to integrable statistics, the data reveal a scale-dependent weakening of rigidity over the experimentally accessible intermediate scales, followed by finite-size-dominated fluctuations at the largest decimation orders.

\begin{table}[htbp]
\centering
\caption{Spectral exponent $\alpha$ from the Theil-Sen fit of the
decimated accumulated PSD.}
\label{tab:alpha}
\begin{tabular}{@{}lcccccccc@{}}
\toprule
$k$ & 1 & 2 & 3 & 4 & 6 & 8 & 10 & 12 \\
\midrule
$^{166}$Er & 1.53 & 1.71 & 1.99 & 1.84 & 2.13 & 2.47 & 1.27 & 1.31 \\
$^{168}$Er & 1.60 & 1.50 & 1.62 & 1.73 & 2.07 & 1.84 & 1.08 & 1.69 \\
\bottomrule
\end{tabular}
\end{table}


\section{Stochastic Hierarchical Framework and Its Application to Erbium Resonances}
\label{sec:htheory_erbium}

\begin{figure*}[t!]
\includegraphics[width=\linewidth]{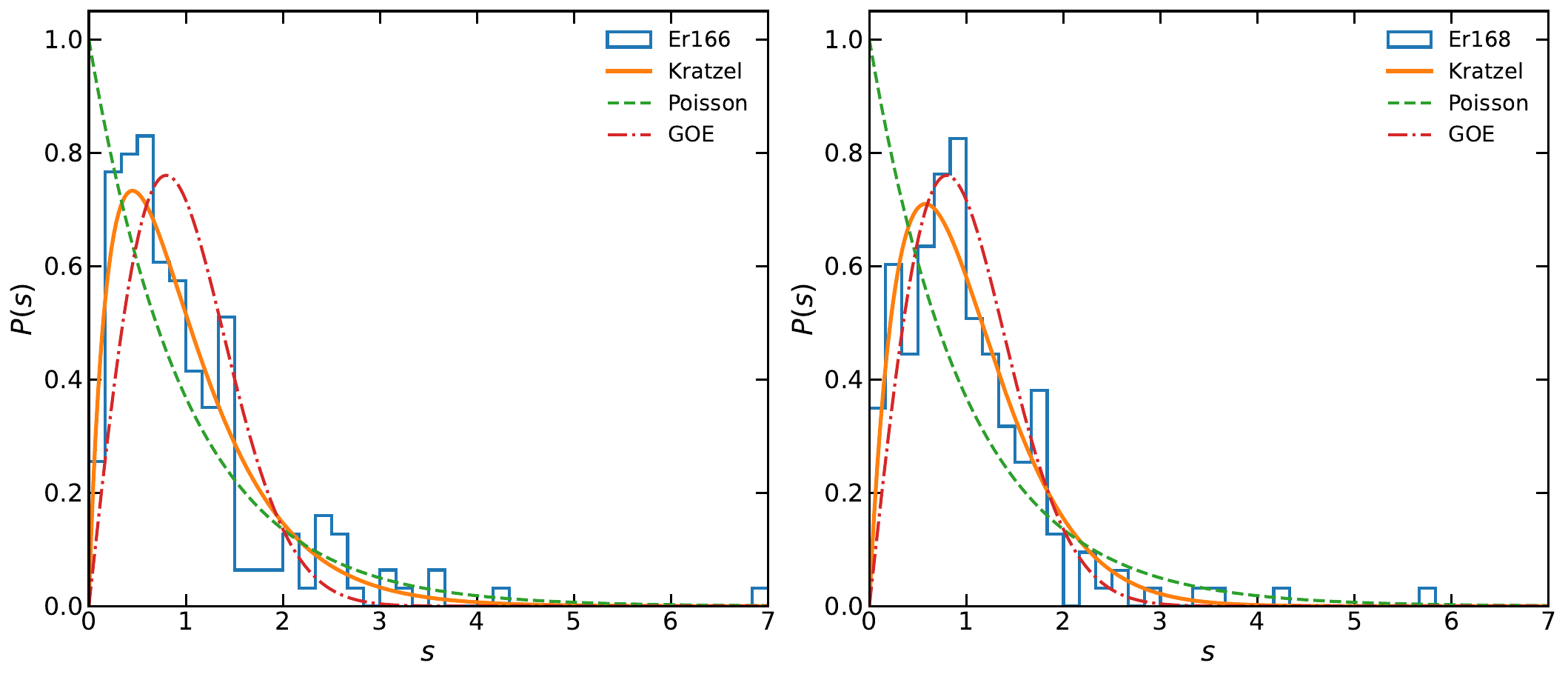}
\caption{Adjacent spacing distributions $P(s)$ with Kr\"atzel fits for
$^{166}$Er (left) and $^{168}$Er (right).  The fitted $(\eta,h)$
parameters indicate linear repulsion with  tails broader than Gaussian tails,
consistent with semi-Poisson affinity.}
\label{fig:kratzel}
\end{figure*}

Although standard random-matrix ensembles successfully describe the asymptotic regimes of quantum dynamics, they do not explicitly incorporate the internal dynamical degrees of freedom associated with continuous crossovers in systems with mixed dynamics. To place the empirical features observed in the erbium Fano-Feshbach spectra  in a phenomenological theoretical context, we adopt the stochastic hierarchical framework, or H-theory. In this approach, the multiscale trapping dynamics characteristic of mixed phase spaces—namely, chaotic regions interwoven with hierarchically organized regular islands—are represented by coupled stochastic differential equations (SDEs) subject to a pronounced separation of time scales \cite{Salazar2010, Macedo2017, Gonzalez2025}.

Within H-theory, the local level spacing $s(t)$ is treated as a fast stochastic degree of freedom governed by a generalized Dyson Brownian motion~\cite{Dyson1962} whose construction is based on the Wigner surmise~\cite{Wigner1951}. By contrast, the local spacing-scale parameter $\varepsilon(t)$ and the relaxation rate $\zeta(t)$ vary slowly and act as fluctuating background fields. Their coupled evolution is described by
\begin{align}
    ds &= - \zeta \left( s - \frac{2 \varepsilon}{s} \right) dt
    + 2 \sqrt{\zeta \varepsilon} \, dW_1,
    \label{eq:sde_s} \\
    d\varepsilon &= -\rho \left(
    \frac{\varepsilon^h}{\varepsilon_0^{h-1}} - \varepsilon_0
    \right) dt
    + \kappa \sqrt{\varepsilon_0 \varepsilon} \, dW_2,
    \label{eq:sde_eps} \\
    d\zeta &= -\tilde{\rho} \left( \zeta - \zeta_0 \right) dt
    + \tilde{\kappa} \sqrt{\zeta_0 \zeta} \, dW_3,
    \label{eq:sde_zeta}
\end{align}
where $W_1$, $W_2$, and $W_3$ denote independent Wiener processes. Equation~\eqref{eq:sde_s} corresponds to a Rayleigh process for the level spacing and therefore enforces local level repulsion. The remaining equations describe slowly varying mean-reverting background fields. In particular, the coefficients $\eta=2\rho/\kappa^2$ and $h$ entering the dynamics of $\varepsilon$ introduce a stochastic modulation of the level-repulsion term, thereby providing a phenomenological parametrization of the coexistence between hierarchical regular islands and a chaotic sea \cite{Zaslavsky2000,Ketzmerick2000}.

A central assumption of the theory is that the background variables evolve on time scales much longer than that of the local level spacing. More explicitly,
\[
\tau_\varepsilon \sim \rho^{-1},
\qquad
\tau_\zeta \sim \tilde{\rho}^{-1},
\qquad
\tau_s \sim \zeta_0^{-1},
\]
with $\tau_\varepsilon,\tau_\zeta \gg \tau_s$. Under this time-scale separation, the joint stationary probability density can be approximately factorized.  The nearest-neighbor spacing distribution and the macroscopic correlation function can then be written as compounded averages over the slow background variables,
\begin{align}
    P(s) &= \int_0^\infty P(s|\varepsilon) f(\varepsilon)
    \, d\varepsilon,
    \label{eq:compound_Ps} \\
    C(\tau) &= \varepsilon_0 \int_0^\infty C(\tau|\zeta) g(\zeta)
    \, d\zeta,
    \label{eq:compound_Ctau}
\end{align}
where $P(s|\varepsilon)$ denotes the conditional Rayleigh distribution derived from the Wigner surmise, whereas $f(\varepsilon)$ and $g(\zeta)$ denote the stationary distributions obtained from the Fokker-Planck equations associated with Eqs.~\eqref{eq:sde_eps} and~\eqref{eq:sde_zeta}, respectively.

The stationary distribution of the  spacing-scale parameter $\varepsilon$ takes the form of a generalized gamma distribution,
\begin{equation}
    f(\varepsilon)
    =
    \frac{h}{\Gamma(\eta/h)}
    \left(\frac{\eta}{h}\right)^{\eta/h}
    \frac{\varepsilon^{\eta-1}}{\varepsilon_0^\eta}
    \exp\left[
    -\frac{\eta}{h}
    \left(\frac{\varepsilon}{\varepsilon_0}\right)^h
    \right],
\end{equation}
where $\eta$ is the corresponding shape parameter. Here $\varepsilon_0$ is a scale parameter which equals the stationary mean of $\varepsilon$ when $h=1$. Inserting the conditional Wigner surmise,
\[
P(s|\varepsilon)
=
\frac{s}{2\varepsilon}
\exp\left(-\frac{s^2}{4\varepsilon}\right),
\]
into Eq.~\eqref{eq:compound_Ps}, and evaluating the resulting Mellin-Barnes contour integral, leads to a marginal NNSD expressed in terms of the Kr\"atzel function $Z_{h,\nu}(x)$~\cite{Kratzel1994,Gonzalez2025},
\begin{equation}
    P(s)
    =
    \frac{s h}{2 \varepsilon_0 \Gamma(\eta/h)}
    \left( \frac{\eta}{h} \right)^{1/h}
    Z_{h,\eta-1}
    \left[
    \frac{s^2}{4\varepsilon_0}
    \left(\frac{\eta}{h}\right)^{1/h}
    \right].
    \label{eq:kratzel_full}
\end{equation}

This result provides a controlled interpolation between several universal spectral regimes. For $h=1$, the expression reduces to the K-distribution, characterized by a modified exponential tail. The integrable Poisson limit is obtained for $\eta=1/2$, whereas the critical semi-Poisson regime is recovered at $\eta=3/2$. When $h>1$, the distribution develops stretched-exponential tails of the form
\[
P(s) \sim \exp\left(-A s^{2h/(h+1)}\right),
\]
which are characteristic of plasma-model statistics in many-body localization and can also arise in phenomenological models of hierarchical trapping in mixed phase spaces \cite{Serbyn2016}.

Fits of the Kr\"atzel distribution in Eq.~\eqref{eq:kratzel_full} to the adjacent-spacing histograms shown in Fig.~\ref{fig:kratzel} yield $\eta\approx1.25$ for both isotopes. The tail behavior is consistent with $h\approx1.0$, although slightly larger values cannot be excluded because the far tail is particularly sensitive to finite-sample effects. The fitted parameters $(\eta,h)$ place both isotopes in the intermediate regime: the repulsion exponent indicates well-developed linear level repulsion, while the tail remains substantially broader than a Gaussian distribution. The value $\eta\approx1.25$ lies between the Poisson limit, $\eta=0.5$, and the semi-Poisson critical point, $\eta=1.5$. Thus, the Kr\"atzel fits provide an independent one-point characterization of the same intermediate spectral behavior identified by the higher-order ratio analysis of Ref.~\cite{Roy2017}. The coexistence of strong local level repulsion with a broad spacing tail is consistent with a spectrum that retains chaotic short-range correlations while exhibiting reduced rigidity at larger scales, as expected for dynamics influenced by hierarchical trapping near regular structures.

\subsection{Hierarchical Correlations and $1/f^\alpha$ Colored Noise}

Whereas the NNSD primarily probes short-range level repulsion, long-range spectral rigidity is controlled by the relaxation-rate variable $\zeta$. Its stationary distribution is a gamma distribution,
\begin{equation}
    g(\zeta)
    =
    \frac{\tilde{\eta}^{\tilde{\eta}}}
    {\Gamma(\tilde{\eta})}
    \frac{\zeta^{\tilde{\eta}-1}}
    {\zeta_0^{\tilde{\eta}}}
    \exp\left(
    -\frac{\tilde{\eta}\zeta}{\zeta_0}
    \right),
\end{equation}
where
\[
\tilde{\eta}
=
\frac{2\tilde{\rho}}{\tilde{\kappa}^2}.
\]
For a fixed value of $\zeta$, the correlation function of the Rayleigh process decays exponentially,
\[
C(\tau|\zeta) \propto \exp(-2\zeta\tau).
\]
However, averaging this exponential relaxation over the broad distribution $g(\zeta)$ in Eq.~\eqref{eq:compound_Ctau} produces an algebraic long-time tail,
\begin{equation}
    C(\tau)
    \propto
    \left(
    \frac{\tilde{\eta}}{\zeta_0}
    + 2\tau
    \right)^{-\tilde{\eta}}
    \sim \tau^{-\tilde{\eta}},
    \qquad
    \tau \gg \zeta_0^{-1}.
\end{equation}

Using the Wiener-Khinchin theorem, together with a regularization of the Fourier transform based on the Gaussian-free-field formulation for $1 \leq \tilde{\eta} \leq 2$, the power spectral density (PSD) of the spectral fluctuations becomes
\begin{equation}
    S(f)
    \sim
    \frac{1}{f^\alpha},
    \qquad
    \text{with}
    \qquad
    \alpha = 3-\tilde{\eta}.
    \label{eq:psd_alpha}
\end{equation}
This relation establishes a direct analytical connection between the stochastic parameters of H-theory and the noise exponents associated with quantum-chaotic spectra. The fully chaotic Wigner-Dyson limit, characterized by $1/f$ noise, is recovered when $\tilde{\eta}\to2$, corresponding to $\alpha=1$. Conversely, the integrable Poisson limit, associated with Brownian $1/f^2$ noise, follows from $\tilde{\eta}\to1$, or equivalently $\alpha=2$.

\begin{figure*}[t]
\includegraphics[width=\linewidth]{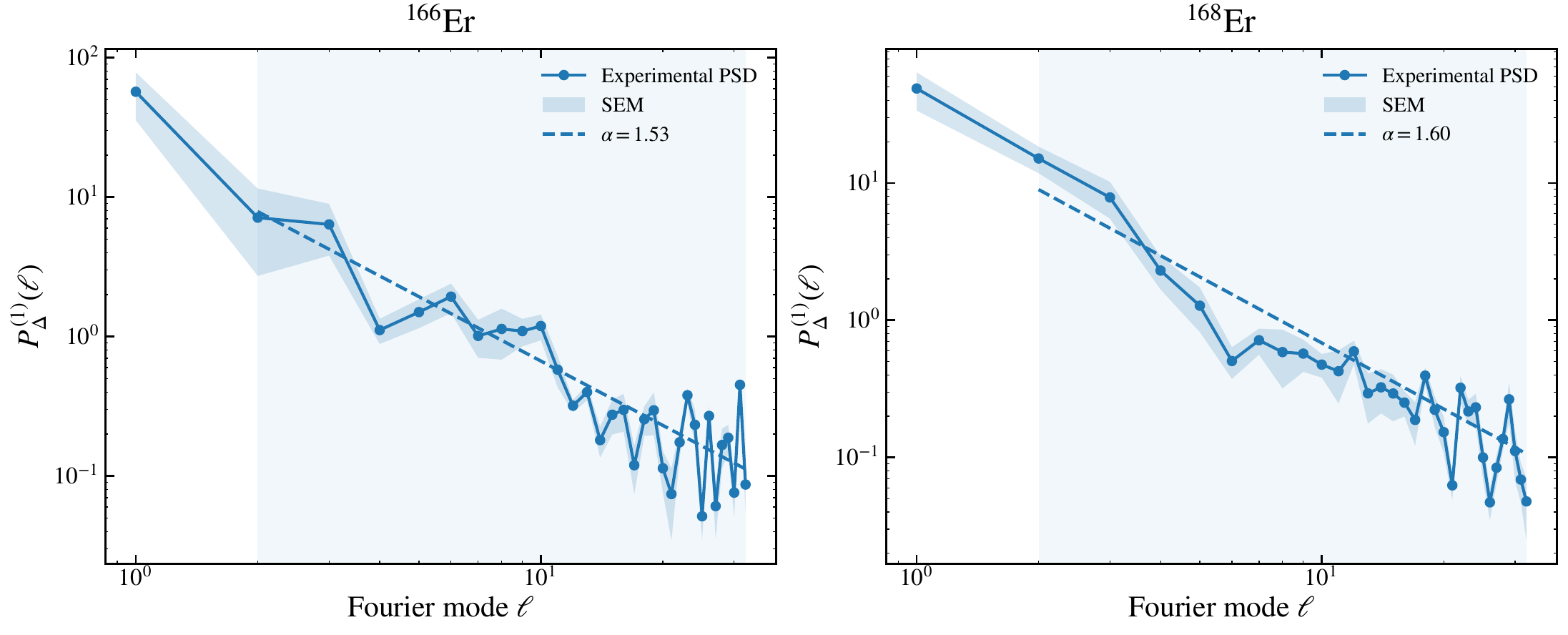}
\caption{
Accumulated power spectral density (PSD), $P^{(1)}_\Delta(\ell)$, for $k=1$, computed using a window length of $L=64$, with 8 records and 32 Fourier modes. The shaded region represents the standard error of the mean (SEM), calculated from the dispersion of the PSD across the individual records and providing an estimate of the uncertainty in the averaged spectrum. Dashed lines indicate the corresponding Theil-Sen power-law fits. The extracted spectral exponents are $\alpha=1.53$ for $^{166}\mathrm{Er}$ and $\alpha=1.60$ for $^{168}\mathrm{Er}$.
}
\label{fig:psd_nnsd}
\end{figure*}

Figure~\ref{fig:psd_nnsd} shows the $k=1$ accumulated PSD in log-log scale.  The Theil-Sen fit over $2\le\ell\le 32$ yields $\alpha=1.53$ for $^{166}$Er and $\alpha=1.60$ for $^{168}$Er. The erbium spectra exhibit $1/f^\alpha$ colored noise that is markedly redder than the $1/f$ behavior expected for fully chaotic systems, indicating reduced spectral rigidity and persistent long-range correlations. Within the hierarchical framework, this behavior is consistent with long-lived memory generated by fluctuating relaxation dynamics across multiple dynamical scales.

H-theory also offers a phenomenological interpretation of the drift of the exponent $\alpha$ observed at higher decimation orders $k$ in Fig.~\ref{fig:psd_grid}. Increasing $k$ causes the decimated accumulated PSD to probe progressively larger mesoscopic spectral scales. From the stochastic perspective, this coarse-graining suppresses the fast fluctuations associated with the chaotic sea and enhances the contribution of  broadly distributed trapping times associated with deeper levels of the island hierarchy. As a result, the effective distribution of relaxation rates can evolve toward $\tilde{\eta}\to1$, for which H-theory predicts a crossover toward the integrable random-walk behavior $\alpha=2$. The observed increase of $\alpha$ toward this regime at intermediate decimation orders is consistent with this picture, while deviations and values above $\alpha=2$ at larger $k$ are naturally subject to finite-size fluctuations in the short decimated sequences. Thus, the scale-dependent weakening of spectral rigidity provides a frequency-domain signature consistent with hierarchical mixed-phase-space dynamics. Together with the higher-order ratio and spacing-increment results,  this supports the interpretation that the erbium resonances are compatible with a multiscale stochastic description of intermediate chaoticity.

\section{Spacing Increments and the Integrable-Chaotic Crossover}

\begin{figure*}[t!]
\includegraphics[width=\linewidth]{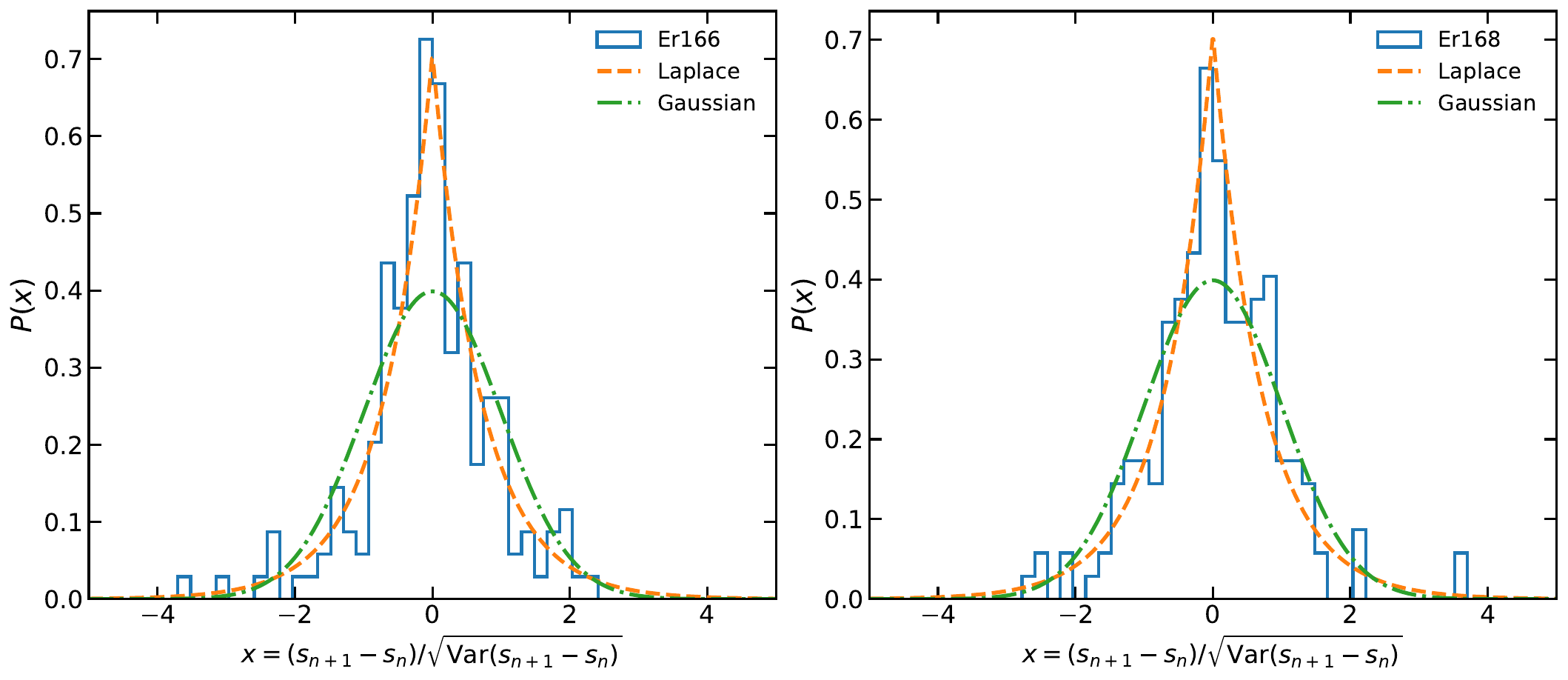}
\caption{Normalized spacing increments $x$ for $^{166}$Er (left) and
$^{168}$Er (right).  Dashed: Laplace (integrable).  Dash-dotted:
Gaussian (chaotic).  The data lie between the two references.}
\label{fig:increments}
\end{figure*}

While the distribution of spacing ratios probes the multiplicative structure of spectral fluctuations, additive spacing increments provide a complementary diagnostic  that is sensitive to local fluctuations of the spectral staircase.  In the study of complex signals, ranging from turbulent flows to mathematical finance, increment statistics provide  a useful characterization of fluctuations across different scales. We define the raw nearest-neighbor spacing increment (NNSI) as $(\delta s)_i \equiv s_{i+1} - s_i$, where $s_i = E_{i+1} - E_i$ are the unfolded nearest-neighbor spacings. To ensure scale invariance and allow for direct comparison across different spectral statistics, we construct the normalized, zero-mean, unit-variance variable:

\begin{equation}
    x_i \equiv \frac{(\delta s)_i}{\sqrt{\operatorname{var}(\delta s)_i}}. 
\end{equation}
Because $(\delta s)_i = E_{i+2} - 2E_{i+1} + E_i$ is a discrete second difference of the ordered level sequence, $P(x)$ probes the local curvature of the spectral staircase rather than its multiplicative structure. Unlike the ratio $r_n$, the increment is not scale invariant. In numerical applications one therefore either unfolds the spectrum or restricts the analysis to a sufficiently narrow bulk window and rescales by the local mean spacing, as done here. After centering and unit-variance normalization, $P(x)$ becomes a symmetric probe of the local anticorrelations between adjacent spacings. This sensitivity is made explicit by the variance decomposition
\begin{equation}
    \operatorname{var}((\delta s)_n) = 2\operatorname{var}(s_n)
    - 2\operatorname{cov}(s_n, s_{n+1}).
    \label{eq:var_increment}
\end{equation}
For Poisson spacings the covariance vanishes and Eq.~\eqref{eq:var_increment} reduces to the standard deviation of the raw Laplace increment distribution
$\sigma_1=\sqrt{2}$. For Wigner-Dyson spectra, in contrast, adjacent gaps are locally anticorrelated, $\operatorname{cov}(s_n,s_{n+1})<0$, consistent with  spectral rigidity: a large gap tends to be compensated by neighboring smaller gaps. 
The anticorrelation enlarges the increment variance relative to the independent-spacing estimate and reshapes the distribution: beta-Hermite simulations show that the normalized Wigner-Dyson increment distributions have a nearly Gaussian central  region, with excess kurtosis close to zero~\cite{Dumitriu2002}, in sharp contrast with the Laplace value of $3$ attained in the regular-integrable limit.

The probability density of the normalized increment, $P(x)$, is fundamentally determined by the joint distribution of two consecutive spacings, $P(s_1, s_2)$. This formalism establishes two universal fixed points that bracket the regular-to-chaotic transition.

In the integrable limit, energy levels are uncorrelated, and the spacings are independent exponential random variables such that $P_I(s_1, s_2) = P_I(s_1)P_I(s_2) = e^{-s_1}e^{-s_2}$. The characteristic function of the normalized increment $x = (s_2 - s_1)/\sigma_1$, with $\sigma_1 = \sqrt{2}$, is readily evaluated as:

\begin{align*}
    \tilde{P}_I(k) = \int_0^\infty ds_1 \int_0^\infty ds_2 \, e^{-s_1 - s_2 - ik (s_2 - s_1)/\sigma_1} \\ = \frac{1}{1 + (k/\sigma_1)^2}.
\end{align*}


 The  law derived above is the exact increment statistics at the integrable fixed point. The corresponding exact result at the chaotic fixed point follows from the same two-gap density. In the bulk GUE limit the spectrum is a determinantal point process, and the joint density of the two consecutive spacings adjacent to a conditioned eigenvalue can be written in terms of a one-point J\'anossy density, $P_c(a_1,a_2) = -\partial_{a_1}\partial_{a_2} J_1(0;[a_1,a_2])$ ~\cite{nishigaki2024}. Fixing one eigenvalue at the origin by Palm conditioning and letting $a_1<0<a_2$ denote its two neighbors, the consecutive spacings are $s_-=-a_1$ and $s_+=a_2$, and the local increment reads $X = s_+ - s_- = a_1 + a_2$. The increment distribution is then the projection
\begin{equation}
    P_X(x) = \frac{1}{2}\int_{|x|}^{\infty} du\,
    P_c\!\left(\frac{x-u}{2}, \frac{x+u}{2}\right),
    \label{eq:PX_projection}
\end{equation}
which makes the symmetry $P_X(x)=P_X(-x)$ manifest whenever $P_c$ is left-right symmetric. Combining Eq.~\eqref{eq:PX_projection} with the J\'anossy representation of $P_c$ yields the exact GUE formula
\begin{equation}
    P_X(x) = -\frac{1}{2}\int_{|x|}^{\infty} du\,
    \frac{\partial^2}{\partial a_1\,\partial a_2}
    J_1(0;[a_1,a_2])
    \bigg|_{\substack{a_1=(x-u)/2\\ a_2=(x+u)/2}}.
    \label{eq:PX_exact_GUE}
\end{equation}
Equation~\eqref{eq:PX_exact_GUE} is the additive counterpart of the exact ratio representation discussed in Sec.~\ref{sec:atas}: once the two-endpoint Tracy-Widom system for the J\'anossy determinant is solved, the same two-gap density generates simultaneously the ratio and the increment statistics. In practice,  while an analytical closed-form expression in terms of elementary functions is not known, the numerical evaluation of Eq.~\eqref{eq:PX_exact_GUE} is computationally tractable. It can be evaluated either by computing the Fredholm determinant with a Nystr\"om or Bornemann-type quadrature~\cite{Bornemann2010},  which provides rapid convergence for sufficiently smooth kernels, and differentiating the resulting function of the endpoints, or by exploiting the integrable form of the transformed kernel, 
$\tilde{K}(\xi,\eta) = [\phi(\xi)\psi(\eta)-\psi(\xi)\phi(\eta)]/(\xi-\eta)$,
for which the determinant and its endpoint derivatives satisfy a closed differential system.
Consequently, Eq.~\eqref{eq:PX_exact_GUE} serves as an accessible benchmark for numerical validation.
For $\beta=1,4$ an analogous construction exists in terms of Pfaffian J\'anossy densities but is algebraically more involved. Eq.~\eqref{eq:PX_exact_GUE} therefore provides the cleanest exact benchmark of the chaotic limit for $P(x)$.


Beyond the regular regime, the onset of quantum chaotic behavior can be modeled as an energy diffusion process described by a probability density function $P(x, T)$ with a parameter $T \ge 1$ that plays the role of a fictitious diffusion time. In this framework,  $T=1$ corresponds to the integrable point with the Laplace distribution, and the evolution of $P(x, T)$ with increasing $T$  phenomenologically represents a mixed phase space, in which islands of stability coexist with an outer domain known as the stochastic sea. As $T$ increases, the model approaches Gaussian statistics, representing the  fully chaotic limit.  

To formalize this transition, we consider a set of continuous random variables $X_j$ with $j=1,\dots,N$. The marginal probability density function is constructed as a convolution sequence~\cite{Feller1971} of the increment distributions for independent steps,  $P(\xi_j) \equiv P(x_{j+1} - x_j)$:
\begin{equation}
    P_N(x_N) = \int \cdots \int \left[ \prod_{j=0}^{N-1} P(x_{j+1} - x_j) dx_j \right] P_0(x_0),
\end{equation}
where $P_0(x_0) = \delta(x_0 - a)$ sets the initial condition. Expressing $P(\xi_j)$ via its Fourier transform,
\begin{equation}
    P(\xi_j) = \int_{-\infty}^\infty \frac{dk}{2\pi} e^{ik\xi_j} e^{\psi(k)},
\end{equation}
and substituting this into the convolution integral yields the characteristic representation of the marginal distribution:
\begin{equation}
    P_N(x_N) = \int_{-\infty}^\infty \frac{dk}{2\pi} e^{ik(x_N - a)} e^{N\psi(k)}.
\end{equation}
Focusing on the determination of the structural function $e^{\psi(k)}$, we apply the inverse Fourier transform.  Taking $P(\xi_j)$  to be the  Laplace increment distribution of the integrable reference process, we obtain
\begin{equation}
    e^{\psi(k)} = \int_{-\infty}^\infty d\xi_j \, e^{-ik\xi_j} P(\xi_j) = \frac{1}{1 + (k/\sigma)^2}.
\end{equation}
Thus, the continuous convolution model for the crossover is determined by the integral
\begin{equation}
    P(x, T) = \int_{-\infty}^\infty \frac{dk}{2\pi} e^{ikx} \left( 1 + \left(\frac{k}{\sigma_T}\right)^2 \right)^{-T},
\end{equation}
where the scale parameter evolves as $\sigma_T = \sqrt{2T}$. This integral admits an exact closed-form representation in terms of the Bessel K-distribution:
\begin{equation}
P(x,T)
=
\frac{2}{\sqrt{\pi}\,\Gamma(T)}
\left(\frac{T}{2}\right)^{\frac{T}{2}+\frac{1}{4}}
|x|^{T-\frac{1}{2}}
K_{T-\frac{1}{2}}\!\left(\sqrt{2T}\,|x|\right).
\end{equation}
This analytical expression interpolates smoothly between the Laplace distribution with zero mean and unit variance at  $P_{T=1}(x) = \frac{1}{\sqrt{2}} \exp\left(-\sqrt{2}\,|x|\right)$, and the Gaussian distribution with zero mean and unit variance as  $P_{T \to \infty}(x) = \frac{1}{\sqrt{2\pi}} \exp\left(-\frac{x^2}{2}\right).$ 

%
For the comparison with finite experimental spectra, the chaotic
reference used in Fig.~\ref{fig:increments} can be sharpened
quantitatively.  For
$X_n = s_{n+1}-s_n$, the increment variance can be written in terms of nearest-spacing moments and the lag-1 spacing correlation,
\begin{equation}
    \sigma_X^2 = 2\left(\langle s^2\rangle - 1\right) - 2 C_1,
    \qquad
    C_1 = \langle s_n s_{n+1}\rangle - 1,
    \label{eq:sigmaX}
\end{equation}
so that $P_X(x)$ is sensitive not only to the one-spacing distribution $P(s)$ but also to the local anticorrelation between neighboring spacings. In Wigner-Dyson spectra $C_1<0$, and the true width exceeds the independent-spacing estimate. Neglecting $C_1$ and inserting the Wigner-Dyson surmise
$P_\beta^W(s) = A_\beta s^\beta e^{-B_\beta s^2}$, with $B_\beta = \left[\Gamma\!\left(\frac{\beta+2}{2}\right)\big/ \Gamma\!\left(\frac{\beta+1}{2}\right)\right]^2$ and $\langle s\rangle=1$, gives the closed independent-spacing Gaussian width
\begin{equation}
    \sigma_{X,\mathrm{ind}}^2(\beta)
    = 2\left[
    \frac{\Gamma\!\left(\frac{\beta+3}{2}\right)
          \Gamma\!\left(\frac{\beta+1}{2}\right)}
         {\Gamma^2\!\left(\frac{\beta+2}{2}\right)} - 1
    \right],
    \label{eq:sigma_ind}
\end{equation}
which evaluates to $\sigma_{X,\mathrm{ind}}^2 \simeq 0.5465$ (GOE), $0.3562$ (GUE), and $0.2089$ (GSE) for unfolded spectra with unit mean spacing. The corresponding Gaussian estimate,
\begin{equation}
    P_{X,\mathrm{ind}}^G(x;\beta)
    = \frac{1}{\sqrt{2\pi\sigma_{X,\mathrm{ind}}^2(\beta)}}
    \exp\!\left[-\frac{x^2}{2\sigma_{X,\mathrm{ind}}^2(\beta)}\right],
    \label{eq:PX_gauss_ind}
\end{equation}
provides a reference central profile with the independent-spacing variance but does not reproduce the exact variance or tails of correlated Wigner-Dyson increment distributions. Systematic improvements are obtained by replacing $\sigma_{X,\mathrm{ind}}^2$ with the exact or numerically computed variance, or by retaining the fourth cumulant $\kappa_4$ in the first Edgeworth correction,
\begin{equation}
    P_X(x) \simeq \frac{e^{-z^2/2}}{\sqrt{2\pi}\,\sigma_X}
    \left[1 + \frac{\kappa_4}{24\,\sigma_X^4} H_4(z)\right],
    \label{eq:edgeworth}
\end{equation}
where $z = \frac{x}{\sigma_X},$  and  $H_4(z) = z^4 - 6z^2 + 3.$ 
After the unit-variance normalization applied to the erbium data,
 the Gaussian approximation reduces to the standard normal law used as the chaotic reference in Fig.~\ref{fig:increments}. Higher-cumulant corrections survive this normalization, so the exact Wigner-Dyson increment density need not be Gaussian. 


In the computational analysis of the experimental erbium spectra, the raw increments are extracted directly from the unfolded Fano-Feshbach resonance sequences of $^{166}$Er and $^{168}$Er. The sequence of differences $\{s_{i+1} - s_i\}$ is centered by subtracting its empirical mean and subsequently divided by its standard deviation to enforce unit variance.

Figure~\ref{fig:increments} shows the normalized spacing-increment distributions obtained from the unfolded Fano-Feshbach resonance spectra of $^{166}$Er and $^{168}$Er, together with the Laplace and Gaussian reference distributions. The two isotopes exhibit very similar behavior. In both cases, the empirical distributions have a rounded central region rather than the cusp characteristic of the Laplace distribution, while their tails extend substantially beyond the Gaussian prediction. Thus, the increment statistics do not coincide with either of the limiting distributions associated with the integrable and chaotic regimes.

The simultaneous broadening of the tails and smoothing of the central peak is characteristic of an intermediate increment distribution. In particular, the enhanced probability of relatively large spacing increments indicates stronger fluctuations than expected for a Gaussian process, whereas the absence of the sharp Laplace cusp distinguishes the spectra from the purely integrable limit. The close agreement between the two erbium isotopes further suggests that this behavior is not specific to a particular isotope, but instead reflects a common statistical property of their resonance spectra.

This result provides an independent characterization of the intermediate spectral regime identified from the higher-order spacing ratios. The ratio statistics probe correlations between levels separated by progressively larger numbers of spacings, whereas the spacing increments quantify the fluctuations between consecutive spacings. The observation of intermediate increment statistics therefore shows that the non-Wigner-Dyson character is also manifested in the local fluctuations of the spectral staircase. Together with the persistence of semi-Poisson-like higher-order ratios, this supports a picture in which strong local level repulsion coexists with enhanced spectral fluctuations and reduced rigidity relative to the fully chaotic random-matrix limit.

\section{Discussion}
\label{sec:discussion}

\subsection{Triangulation of the universality class}

The five diagnostic figures presented in this work do not provide isolated characterizations of the spectral fluctuations. Rather, they offer complementary and mutually consistent views of the same intermediate spectral regime. The Kr\"atzel fit to the adjacent-spacing distribution (Fig.~3) yields a shape parameter $\eta\approx1.25$ with a tail-shape exponent $1<h<2$, indicating pronounced local level repulsion together with broad, non-Gaussian spacing fluctuations. This one-point characterization is consistent with the higher-order spacing-ratio analysis (Fig.~1), where the experimental distributions exhibit a clear affinity for the semi-Poisson family at $k=1$ and remain systematically broader than the effective-index Wigner-Dyson predictions at larger decimation orders.

Moving to the frequency domain, the $k=1$ accumulated power spectral density (Fig.~4) yields an exponent $\alpha\approx1.5$-$1.6$, lying between the chaotic $1/f$ and integrable $1/f^2$ limits. Crucially, the hierarchy of decimated PSDs (Fig.~2) shows that this intermediate rigidity is scale dependent: at intermediate decimation orders, the exponent moves toward the random-walk limit $\alpha=2$, indicating a weakening of spectral rigidity at mesoscopic scales, while the largest decimation orders become increasingly affected by finite-size fluctuations. Finally, the spacing-increment distributions (Fig.~5) provide an independent characterization of the integrable-chaotic crossover, exhibiting statistics between the Laplace and Gaussian limiting distributions and corresponding to a finite crossover parameter $T$.

Taken together, these complementary observables establish a consistent picture of the erbium Fano-Feshbach spectra as an intermediate, non-Wigner-Dyson regime. Strong local level repulsion coexists with weaker higher-order correlations, reduced long-range rigidity, and broad spacing fluctuations. The agreement among these independent diagnostics strengthens the interpretation that the observed semi-Poisson character is not restricted to nearest-neighbor statistics, but reflects a broader multiscale organization of spectral fluctuations.

\subsection{Connection to previous semi-Poisson analyses}

Our multi-scale findings build directly upon the successive spectral-statistics analyses of the erbium Fano-Feshbach resonance spectra. Frisch \textit{et al.}~\cite{Frisch2014} first established strong signatures of quantum chaos in the $^{166}$Er and $^{168}$Er collision spectra through comparisons with random-matrix predictions. Subsequently, Mur-Petit and Molina~\cite{MurPetit2015} examined both short- and long-range spectral fluctuations and identified systematic deviations from the standard random-matrix description, discussing, in particular, the possible influence of missing resonances. These results indicated that the spectra possess chaotic signatures while retaining statistical features that are not fully captured by conventional random-matrix ensembles.

Roy \textit{et al.}~\cite{Roy2017} subsequently revisited the same experimental datasets using unfolding-independent spacing-ratio statistics. They found that the empirical ratio distributions lie between the Poisson and GOE limits, with the closest agreement obtained with the semi-Poisson ratio law $6r/(1+r)^4$. Their analysis therefore established the semi-Poisson character of the erbium spectra at the nearest-neighbor level.

Our results extend this progression by addressing the scale dependence of these intermediate statistics. The higher-order ratio distributions corroborate the semi-Poisson character at $k=1$ and show that the experimental spectra remain systematically closer to the semi-Poisson family than to the effective-index Wigner-Dyson predictions at larger level separations. In parallel, the accumulated PSD reveals an exponent $\alpha\approx1.5$-$1.6$ at full spectral resolution and a tendency toward the integrable random-walk regime at intermediate decimation orders. This indicates that the spectral rigidity is itself scale dependent and weakens when the spectrum is coarse-grained.

Thus, the present analysis does not simply provide another confirmation of the semi-Poisson statistics reported in Ref.~\cite{Roy2017}. Rather, it connects the earlier observations of deviations from conventional random-matrix behavior~\cite{MurPetit2015} and the subsequent identification of semi-Poisson statistics~\cite{Roy2017} to a multi-scale picture in which intermediate spectral correlations persist beyond nearest neighbors and spectral rigidity decreases at mesoscopic scales. In this sense, the present work provides a broader characterization of the statistical regime underlying the anomalous spectral fluctuations of the erbium resonances.

\subsection{Connection to the integrable-chaotic crossover}
The statistics of the normalized spacing increments (Fig.~5) offer a direct bridge to the stochastic dynamics framework recently developed by Gonz\'alez \textit{et al.}~\cite{Gonzalez2025}. In their formulation, the nearest-neighbor spacing increment (NNSI) density evolves continuously from the Laplace fixed point ($T=1$, representing the integrable limit) to the Gaussian attractor ($T \to \infty$, representing the fully chaotic limit) via a diffusion process characterized by an effective fictitious time $T = 2\eta$.

The empirical erbium data exhibit a finite, intermediate value of the crossover parameter, consistent with the Kr\"atzel shape parameter $\eta\approx1.25$ extracted independently from the adjacent-spacing distributions (Fig.~3). The corresponding value $T\approx2.5$ places the spectra well away from both limiting distributions. This correspondence provides a quantitative connection between the one-spacing Kr\"atzel characterization and the two-spacing increment statistics within the stochastic crossover framework.

The agreement supports the interpretation that the anomalous spacing fluctuations of the erbium resonances can be represented within this framework as an intermediate stochastic regime between the integrable and fully chaotic limits. Importantly, this interpretation complements rather than replaces the spectral evidence: the higher-order ratios and accumulated PSD independently demonstrate persistent non-Wigner-Dyson correlations and scale-dependent spectral rigidity. Together, these results suggest that the intermediate statistics of the erbium resonances can be understood as the spectral manifestation of dynamics lying between the integrable and fully chaotic regimes.

\subsection{Limitations and finite-size effects}

While the multi-scale diagnostics provide a coherent picture of the spectral fluctuations, the finite number of experimentally observed resonances ($N\approx190$) should be kept in mind when interpreting the results at large decimation orders. As $k$ increases, the decimated sequences become shorter and the number of available residue classes is reduced. Consequently, the statistical fluctuations of the accumulated PSD become more pronounced at larger $k$, as reflected in the variation of the extracted exponent $\alpha$. The values obtained at the largest decimation orders should therefore be interpreted as characterizing the spectral behavior over the accessible finite-size range rather than as estimates of an asymptotic limit.

Finite-size considerations also apply to the higher-order ratio distributions $P^{(k)}(r)$. The observed narrowing with increasing $k$ is consistent with the effective-index description, although the narrowing itself does not imply a progressively stronger approach to Wigner-Dyson statistics. The comparison with the corresponding effective-index predictions is therefore important for assessing the nature of the correlations at larger level separations.

Overall, the finite size of the available spectra limits the range of scales over which asymptotic behavior can be established quantitatively. Importantly, however, the main observations of the present work remain unchanged over the experimentally accessible range: the higher-order ratio distributions retain their non-Wigner-Dyson character, while the accumulated PSD exhibits an intermediate exponent at full spectral resolution and a tendency toward weaker rigidity at intermediate decimation scales. The finite-size effects primarily become more pronounced at the largest decimation orders, where the statistical resolution is reduced. Further measurements with longer resonance sequences would allow these trends to be tested over a broader range of spectral scales.

\section{Conclusions}
\label{sec:conclusions}

In summary, we have presented a comprehensive multi-scale statistical analysis of the Fano-Feshbach resonance spectra of ultracold $^{166}$Er and $^{168}$Er atoms. By combining three complementary diagnostics--higher-order spacing ratios, normalized spacing increments, and the decimated accumulated power spectral density--we have characterized the spectral regime of this complex atomic system across multiple scales. Our central finding is that the erbium spectra exhibit intermediate quantum-chaotic statistics: they display robust local level repulsion characteristic of chaotic dynamics, while their higher-order correlations and long-range spectral rigidity remain weaker than expected for standard Wigner-Dyson ensembles. The resulting combination of strong short-range repulsion and reduced spectral rigidity is consistent with the semi-Poisson character previously identified in these spectra.

This analysis corroborates and significantly extends the nearest-neighbor ratio study of Roy \textit{et al.}~\cite{Roy2017}. While their work established the close-to-semi-Poisson character of the erbium spectra at the nearest-neighbor level, our higher-order ratio analysis demonstrates that the non-Wigner-Dyson character persists across larger level separations. The normalized spacing-increment statistics provide a complementary connection to the stochastic integrable-chaotic crossover framework of reference~\cite{Gonzalez2025}. Within this framework, the crossover parameter is related to the Kr\"atzel shape parameter through $T=2\eta$, giving $T\approx2.5$ for the fitted value $\eta\approx1.25$. This correspondence provides a quantitative connection between the adjacent-spacing and spacing-increment descriptions of the intermediate regime.

Furthermore, our frequency-domain analysis reveals a scale-dependent weakening of spectral rigidity. At full spectral resolution, the accumulated PSD yields $\alpha\approx1.5$-$1.6$, intermediate between the chaotic $1/f$ and integrable $1/f^2$ limits. At intermediate decimation orders, the exponent moves toward the random-walk regime, indicating that spectral rigidity decreases upon coarse-graining over larger level separations. At the largest decimation orders, finite-size effects become substantial and limit the interpretation of the extracted exponents. Thus, the principal result is not a monotonic flow toward $\alpha=2$, but the observation of an intermediate-scale reduction of spectral rigidity that complements the persistent non-Wigner-Dyson correlations revealed by the higher-order ratios.

Taken together, the higher-order ratios, spacing increments, and accumulated PSD provide a consistent multi-scale characterization of the erbium resonance spectra. The combined evidence supports an interpretation in terms of intermediate quantum chaos and is compatible with a hierarchical mixed-phase-space picture in which strong local level repulsion coexists with reduced spectral rigidity at larger scales. More broadly, the methodology developed here provides a framework for probing the scale dependence of spectral correlations in complex quantum systems where conventional nearest-neighbor statistics alone are insufficient.

An important open question concerns the microscopic dynamical origin of the observed scale-dependent spectral correlations. While the present results are consistent with a hierarchical mixed-phase-space picture in which trapping near regular structures weakens spectral rigidity at mesoscopic scales, the available resonance spectra do not by themselves establish a direct correspondence between specific dynamical structures and the observed spectral exponents. A natural direction for future work is therefore to combine spectral statistics with a direct characterization of the underlying collision dynamics, in order to determine how hierarchical trapping and the associated distribution of dynamical time scales give rise to the observed intermediate spectral regime.

\begin{acknowledgments}
This work was supported in part by the following Brazilian agencies: Conselho Nacional de Desenvolvimento Científico e Tecnológico (CNPq), Coordenação de Aperfeiçoamento de Pessoal de Nível Superior (CAPES), and Fundação de Amparo à Ciência e Tecnologia do Estado de Pernambuco (FACEPE). A.M.S.M. acknowledges financial
support from CNPq (Grant  No.310800/2026-9).
\end{acknowledgments}

\bibliographystyle{apsrev4-2}
\bibliography{ref}

\end{document}